\documentstyle[preprint,aps]{revtex}
\title{Fermion to boson mappings revisited}
\author{Joseph N.~Ginocchio and Calvin W.~Johnson}
\address{T-5, MS B283, Theoretical Division, Los Alamos National Laboratory,
Los Alamos, NM
87545}

\begin{document}
\draft

\maketitle
\begin{abstract}
\noindent
We briefly review various mappings of fermion pairs to
bosons, including those based on mapping operators, such as
Belyaev-Zelevinskii, and those on mapping states, such as Marumori;
in particular we consider the work of Otsuka-Arima-Iachello,
aimed at deriving the Interacting Boson Model.
We then give a rigorous and unified
description of state-mapping procedures which allows one to systematically
go beyond Otsuka-Arima-Iachello and related approaches, along with
several exact results.
\end{abstract}

\section{ Introduction}

Professor Belyaev (with V. G. Zelevinskii) (BZ)
 pioneered the mapping of fermion
systems onto bosons more than thirty years ago \cite{BZ}.
These original attempts at bosonization of fermion systems was
motivated by collective particle-hole modes in nuclei.  Since
that time the interacting boson model \cite{IBM} (IBM) has been
phenomenologically very successful in explaining low energy nuclear
spectroscopy for heavy nuclei.
The bosons in this model are thought to represent
monopole (J=0), quadrupole (J=2), and sometimes hexadecapole (J=4)
correlated pairs of valence nucleons in the shell model.
The IBM
Hamiltonian is Hermitian, usually has at most two-boson interactions,
and conserves boson number, reflecting the particle-particle, rather than
particle-hole, nature of the underlying fermion pairs.
While one can numerically
diagonalize the general IBM Hamiltonian, one of the strengths of IBM are
the algebraic limits corresponding to the subgroups SU(3), U(5), or O(6),
with analytic expressions for excitation bands and
transition strengths, which encompass an enormous amount of
nuclear data.

The microscopic reasons for the success of such a simple model are elusive.
Otsuka, Arima, and Iachello, along with Talmi,
have used a mapping of the shell model Hamiltonian to the
IBM Hamiltonian \cite{OAIT,OAI} based on the
seniority model \cite{Talmi}, but these attempts have not done well for
well-deformed nuclei.  For this reason we have revisited boson
mappings to see if we can understand the success of the IBM starting
from the shell model.

In the next section we sketch out various historic approaches to boson
map\-pings \cite{Review}.  We follow Marumori \cite{Maru} and
Otsuka et al.\ \cite{OAIT,OAI} (OAI)
in our mapping procedure which maps fermion states into boson states and
construct boson operators that reproduce fermion matrix elements.
We give the boson representation of the
 Hamiltonian and review the result that, in the full space, it
factorizes into a boson image, which is the same as the BZ
Hamiltonian, times a normalization operator which projects out the
spurious states.
However, since our goal is to understand the IBM which only
deals with a few of the enormous degrees of freedom of the shell model,
we go on to discuss boson images in truncated spaces.  This, we shall see,
gives rigorous insight into the OAI mapping and shows how to go systematically
beyond it.

\section{A brief history of boson mappings}

The fundamental goal is to solve the many-fermion Schr\"odinger
equation
\begin{equation}
\hat{H} \left | \Psi_\lambda \right \rangle = E_\lambda
 \left | \Psi_\lambda \right \rangle
\label{Schrodinger1}
\end{equation}
and find transition matrix elements between eigenstates,
$t_{\lambda \lambda^\prime} =
\left \langle \Psi_\lambda \left | \hat{T}
 \right | \Psi_{\lambda^\prime} \right \rangle$.    As the fermion Fock
space may be so large as to make direct solution intractable, the idea of
a boson mapping is to replace the fermion operators with boson operators,
{\em using only a minimal number of boson degrees of freedom}, that approximate
the spectrum and transition matrix elements of the original fermion problem.
There are two approaches to boson mappings which we now review.

The first approach, epitomized by
Belyaev and Zelevinskii\cite{BZ},
is to map fermion operators to boson operators
so as to preserve the original algebra. Specifically, consider a
space with $2\Omega$ single-fermion states; $a^\dagger_i, a_j$ signify
fermion creation and annihilation operators.
The set of all bilinear femion operators,
$a_ia_j, a^\dagger_k a^\dagger_l, a^\dagger_i a_j$, form the Lie algebra of
${\rm SO}(4\Omega)$, as embodied by the commutation
relations
\begin{eqnarray}
\left [ a_i a_j, a_k a_l \right ] = 0  \\
\left [ a_i a_j, a^\dagger_k a^\dagger_l \right ] =
\delta_{il} \delta_{jk} +
\delta_{ik} a^\dagger_l a_j
+\delta_{jl} a^\dagger_k a_i
- (i \leftrightarrow j)
 \\
\left [ a_i a_j, a_k^\dagger a_l \right ] =
\delta_{jk} a_i a_l - (i \leftrightarrow j)
 \\
\left [ a^\dagger_i a_j, a^\dagger_k a_l \right ] =
\delta_{jk} a^\dagger_i a_l -\delta_{il} a^\dagger_k a_j
\end{eqnarray}
At this point it is convenient to introduce collective fermion
pair operators
\begin{equation}
\hat{A}^\dagger_\beta \equiv  { 1 \over \sqrt{2}} \sum_{ij}
\left ( {\bf A}^\dagger_\beta \right )_{ij}
{a}^\dagger_i {a}^\dagger_j.
\end{equation}
We always choose the  $\Omega(2\Omega-1)$ matrices ${\bf A}_\beta$ to be
antisymmetric to
preserve the underlying fermion statistics, thus eliminating the need
later on to distinguish between `ideal' and `physical' bosons.
We also assume the following normalization and completeness relations for
the matrices:
\begin{eqnarray}
{\rm tr\,} {\bf A}_\alpha {\bf A}^\dagger_\beta = \delta_{\alpha \beta}; \\
\sum_\alpha \left ({\bf A}^\dagger_\alpha\right )_{ij}
\left ({\bf A}_\alpha\right )_{j^\prime i^\prime}
={1\over 2} ( \delta_{i i^\prime} \delta_{j j^\prime}
-\delta_{i j^\prime}  \delta_{j i^\prime} )  .
\label{completeness}
\end{eqnarray}
Generic one- and two-body
fermion operators we represent by
$ \hat{T} \equiv
\sum_{ij} T_{ij} {a}^\dagger_i {a}_j$,
$\hat{V} \equiv \sum_{\mu \nu} \left \langle \mu
\right | V \left |  \nu \right \rangle
\hat{A}^\dagger_\mu \hat{A}_\nu,$
where $T_{ij} = \left \langle i \right | \hat{T} \left | j \right \rangle$;
from such operators one can construct a fermion Hamiltonian $\hat{H}$.
Now one has the following commutation relations:
\begin{eqnarray}
\left [ \hat{A}_\alpha , \hat{A}_\beta \right ] =
\left [ \hat{A}^\dagger_\alpha , \hat{A}^\dagger_\beta \right ] = 0;
\label{commute1} \\
\left [ \hat{A}_\alpha , \hat{A}^\dagger_\beta \right ] =
\delta_{\alpha \beta} -2 \sum_{ij}
\left ( {\bf A}^\dagger_\beta {\bf A}_\alpha \right )_{ij}
a^\dagger_i a_j ; \\
\left [ \hat{A}_\alpha , \hat{T} \right ] =
2\sum_\beta {\rm tr \,} \left (
{\bf A}_\alpha {\bf T} {\bf A}^\dagger_\beta \right ) \hat{A}_\beta
\\
\left [ \hat{T}_1, \hat{T}_2 \right ]
= \sum_{ij} \left ( {\bf T}_1{\bf T}_2 -{\bf T}_2{\bf T}_1 \right)_{ij}
a^\dagger_i a_j
\label{commutelast}
\end{eqnarray}
The method of Belyaev and Zelevinskii is to find boson images of the
bifermion operators,
\begin{eqnarray}
(\hat{A}^\dagger_\mu)_B = b^\dagger_\mu + \sum_{\alpha \beta \gamma}
x_\mu^{\alpha \beta \gamma} b^\dagger_\alpha b^\dagger_\beta b_\gamma
+ \sum_{ \alpha \beta \gamma \delta \epsilon }
x_\mu^{\alpha \beta \gamma \delta \epsilon }
b^\dagger_\alpha b^\dagger_\beta  b^\dagger_\delta b_\gamma b_\epsilon
+ \ldots
\label{BZpair}
\\
(\hat{A}_\mu)_B = (A^\dagger_\mu)_B^\dagger \\
(\hat{T})_B =
\sum_{\alpha \beta} y^{\alpha \beta} b^\dagger_\alpha b_\beta
+\sum_{\alpha \beta \gamma \delta } y^{\alpha \beta \gamma \delta }
 b^\dagger_\alpha b^\dagger_\gamma b_\beta b_\delta
+ \ldots
\end{eqnarray}
where $b_\alpha, b^\dagger_\beta$ are boson creation and annihilation
operators, $[ b_\alpha, b^\dagger_\beta ] = \delta_{\alpha \beta}$,
with the coefficients $x, y$
chosen so that the images $(A^\dagger_\mu)_B, (A_\nu)_B, (T)_B$
 have the same commutation relations as in
(\ref{commute1})-(\ref{commutelast}).
Because the algebra is exactly matched, if one builds boson states in
exact analogy to the fermion states
then the full boson Fock space is not spanned and one does not have
nonphysical or spurious states.  For the state-mapping methods
described below, especially when the entire boson Fock space is used,
identifying and decoupling from spurious states is an important and
problematic  issue.

 On the other hand, the Belyaev-Zelevinskii
expansion is in general infinite.  In the full boson Fock space, that is,
no truncation of the boson degrees of freedom, the image
of one body operators is finite and given quite simply by
$(\hat{T})_B  = 2 \sum_{\alpha \beta} {\rm tr \,} ({\bf A}_\alpha {\bf T}
{\bf A}^\dagger_\beta ) b^\dagger_\alpha b_\beta$.
Since any fermion Hamiltonian can be written in terms of one-body operators,
the boson image of a finite fermion Hamiltonian can be finite in the full
space.  The states that one must use then are built from the pairs given in
(\ref{BZpair}) which will not just be products of bosons but will include
exchange terms.  For example, for two bosons and using (\ref{BZpair}),
\begin{equation}
\hat{A}^\dagger_\alpha \hat{A}^\dagger_\beta
\left | 0 \right \rangle \rightarrow
\left ( b^\dagger_\alpha b^\dagger_\beta
+ x^{\sigma \tau\beta}_{\alpha}  b^\dagger_\sigma b^\dagger_\tau \right )
\left | 0 \right ).
\end{equation}
These exchange terms are due to the antisymmetry.  We shall take care of
such exchange effects by introducing a norm operator in the boson space.
For truncated spaces, however, the expansion of the BZ Hamiltonian is
infinite.

The Dyson mapping \cite{Dyson} is a variant of Belyaev-Zelevenskii, in
which one makes the mapping
\begin{eqnarray}
\hat{A}_\alpha \rightarrow b_\alpha; \\
\hat{A}^\dagger_\beta \rightarrow b^\dagger_\beta
- 2\sum_{\lambda \mu \nu}
{\rm tr \,} ( {\bf A}_\lambda {\bf A}^\dagger_\beta
{\bf A}_\mu {\bf A}^\dagger_\nu )
b^\dagger_\lambda b^\dagger_\mu b_\nu\\
\hat{T} \rightarrow
\sum_{\alpha \beta}  2 {\rm tr \,} ({\bf A}_\alpha {\bf T}
{\bf A}^\dagger_\beta ) b^\dagger_\alpha b_\beta.
\\
\end{eqnarray}
The operators are then clearly finite; on the other hand they are just as
clearly non-Hermitian.  From a computational viewpoint non-Hermiticity is
only a minor
barrier, but it is an obstacle to an understanding of the microscopic
origin of Hermitian IBM Hamiltonians.  Furthermore the Dyson operators
mix spurious and physical spaces.

Marshalek \cite{Marshalek} points out there exist mappings that
are both finite and Hermitian,
but these in general require projection operators to eliminate
spurious states.  We will regain this result later on in this paper.

The second major approach, pioneered by Marumori \cite{Maru},
 is to map fermion states and construct boson
operators that preserve matrix elements.
For fermion many-body (shell-model)
basis states one often
uses Slater determinants, antisymmetrized products of single-fermion
wavefunctions which we can write using Fock creation operators:
$a^{\dagger}_{i_{1}} \cdot \cdot \cdot
a^{\dagger}_{i_{n}} \left | 0 \right \rangle $ for $n$ fermions.
For an even number of fermions one instead constructs states from products
of fermion pairs,
\begin{equation}
\left | \Psi_\beta \right \rangle
= \prod_{m=1}^N\hat{A}^\dagger_{\beta_m} \left | 0 \right \rangle;
\label{StateDefn}
\end{equation}
if the number of fermion is fixed at $n$ then $m$ runs from 1 to $N=n/2$.
The original work of Marumori, however, focused on particle-hole excitations
and so the number of pairs and consequently bosons is not fixed.
These states are not trivially orthonormal.  They must be orthogonalized.
Exactly how this orthogonalization is accomplished will be a key theme in
this paper.

Marumori constructs the norm matrix
\begin{equation}
{\cal N}_{\alpha \beta} = \left \langle \Psi_\alpha | \Psi_\beta
\right \rangle
\end{equation}
and then the Usui operator
\begin{equation}
 U = \sum_{\alpha, \beta; n}
\left | \Phi_\beta \right )
\left ( {\cal N}\right)^{-1/2}_{\beta \alpha}
\left \langle \Psi_\alpha \right |
\end{equation}
where
bosons states are constructed in strict
analogy to the fermion states,
\begin{equation}
\left | \Phi \right \rangle =
\prod_{m=1}^N b_{\beta_m} \left | 0 \right \rangle.
\end{equation}
Then the Marumori expansion of any fermion operator is
\begin{equation}
O_B =  U O_F  U^\dagger.
\end{equation}
Clearly Marumori is best suited for particle-hole states with only a
few excitations.  If one applied it to a system with numerous
particle-particle pairs, as for the IBM,  one obtains clumsy many-body
terms.   Kishimoto and Tamura \cite{KT} addressed this
last issued by introducing a ``linked-cluster'' expansion which they then
grafted into a BZ-type scheme.

Otsuka, Arima, and Iachello (OAI), along with Talmi \cite{OAIT,OAI},
investigated the microscopic origins of the Interacting Boson Model
through boson mappings.  Although they also mapped states, they differed
from Marumori in some key details.  First of all, they built states built
on a fixed number of particle-particle, not particle-hole, pairs,
and restricted the pairs to one monopole ($J^\pi = 0^+$) and
quadrupole ($J^\pi = 2^+$) pair.
These states were orthogonalized based on seniority.
That is, they construct, for $2N$ fermions,
 low-seniority basis states of $S$ and $D$ fermion pairs,
$\left | S^{N-n_d} D^{n_d} \right \rangle$, and then orthonormalize the
states such that the zero-seniority state is mapped to itself, and
states of higher seniority $v$ are orthogonalized against states of lower
seniority,
\begin{equation}
\left | v\right ) \rightarrow \left | ``v" \right ) =\left | v \right )
+ \left | v-2 \right ) + \left | v -4 \right ) + \ldots
\end{equation}
Then OAI calculate the matrix elements
$ \left \langle `` S^{N-n_d^\prime} D^{n_d^\prime } "\left |
H_F \right | `` S^{N-n_d} D^{n_d} "\right \rangle$ for
$n_d, n_d^\prime = 0,1,2$ and obtain the coefficients
for their one plus two-boson Hamiltonian. These coefficients have an explicit
$N$-dependence
(and for large $N$ and arbitrary systems such matrix elements
are not trivial to calculate, especially in analytic form!)
and thus implicitly a many-body dependence. At first sight this is not entirely
unreasonable as it is well known the
IBM parameters change substantially as a function of the number of
bosons, even within a major `shell'.
Nonetheless the OAI mapping has three drawbacks. The first is that
it's not clear how to systematically calculate many-body contributions
beyond that contained in the OAI prescription, whereas the  method we shall
describe is fully and rigorously systematic.
The second is that the OAI
prescription can induce many-body effects where none are needed.
This point will be illustrated in
section \ref{ApproxImages}.  Thirdly, only the $n_d = 0,1,2$
space is exactly mapped, but very deformed systems will involve large $n_d$.
In fact, for an axial rotor limit, the average number of d-bosons in the ground
 state band is 2/3 the total number of bosons.

As an alternative to OAI, Skouras, van Isacker, and Nagarajan
\cite{democratic} proposed a
``democratic'' mapping where the orthogonalization is based on
eigenvectors of the norm matrix rather than seniority.

In what follows we attempt to rigorously unify all the state-mapping
methods.  We have three strong results.  First, we give
general expressions for fermion matrix elements via boson representations.
Second, we show how in several cases one can have exact, finite, and
Hermitian boson images of fermion operators.   Finally, we show how to
extend both the OAI and democratic mappings in a systematic and
rigorous fashion, and illustrate how the choice of orthogonalization
can affect the many-body dependence of the boson images.

\section{Boson representations of fermion matrix elements}

The starting point of any state-mapping method is the calculation of
matrix elements of fermion operators between states constructed from
fermion pairs of the form (\ref{StateDefn}), including the overlap:
$\left \langle \Psi_\alpha | \Psi_\beta \right \rangle $,
$\left \langle \Psi_\alpha \left | \hat{H} \right |
 \Psi_\beta \right \rangle $,
$\left \langle \Psi_\alpha \left | \hat{T} \right |
 \Psi_\beta \right \rangle $, and so on.
These matrix elements are much more difficult to compute
 than the corresponding matrix
elements between Slater determinants.  As we shall show, however, full and
careful attention paid to the problem of calculation matrix elements can
yield powerful results.
Silvestre-Brac and Piepenbring \cite{Silvestre},
laboriously using commutation relations, derived a  Wick theorem for
fermion pairs.   Rowe, Song and Chen \cite{Rowe}
using `vector coherent states'
(we would say fermion-pair coherent states) found matrix elements between
pair-condensate wavefunctions, states of the form
$\left ( \hat{A}^\dagger \right )^N \left | 0 \right \rangle.$
Using a theorem by Lang et al.\ \cite{Lang}, we have
generalized \cite{GJ} the method of Rowe, Song and Chen and
recovered (actually
discovered independently) the
expressions of Silvestre-Brac and Piepenbring.
Specifically, we construct  generating functionals by taking the matrix
element
\begin{eqnarray}
\left \langle 0
\left | \exp \left ( \sum_\alpha \epsilon_\alpha \hat{A}_\alpha \right )
 \exp \left ( \sum_\beta \epsilon_\beta \hat{A}_\beta^\dagger \right )
\right | 0 \right \rangle \nonumber \\
= \exp \left ( \sum_{k = 1}^\infty
{ (-2)^{k-1} \over k } {\rm tr \,} \left [
\sum_{\alpha \beta} \epsilon_\alpha \epsilon_\beta
{\bf A}_\alpha {\bf A}^\dagger_\beta \right ]^k \right ) .
\label{generating}
\end{eqnarray}
By taking derivatives of $\epsilon_\alpha$, etc., one computes the
desired matrix elements in analytic form\cite{GJ}.
 For pair condensate wave functions
one can calculate the matrix elements iteratively and propose a
variational principle \cite{Rowe}.
Such a variational principle would be useful in determining the
``best'' microscopic structure for a truncated set of pairs.
Alternately,  Otsuka and Yoshinaga \cite{OY}
choose their $S$ and $D$ pairs from Hartree-Fock-Bogoliubov states; the
two approaches can probably be related in some approximation.
This may be important is answering a basic question of IBM, the
origin of algebraic limits: do they arise from changes in pair structure,
or from effective many-body effects, or both?

We now want to translate the fermion matrix elements into boson space.
We take the simple mapping of fermion states into boson
states
\begin{equation}
 \left | \Psi_\beta \right \rangle
\rightarrow  \left | \Phi_\beta \right )
= \prod_{m=1}^N {b}^\dagger_{\beta_m} \left | 0 \right ),
\end{equation}
where the ${b}^\dagger$ are boson creation operators.
We construct
 boson operators that preserve matrix elements, introducing boson operators
$\hat{\cal T}_B$, $\hat{\cal V}_B$, and most
importantly the {\it  norm operator} $\hat{\cal N}_B$
 such that
$ \left ( \Phi_\alpha \right | \hat{\cal T}_B \left | \Phi_\beta \right )
= \left \langle \Psi_\alpha \right| \hat{T} \left | \Psi_\beta \right \rangle$,
$
\left ( \Phi_\alpha \right | \hat{\cal V}_B \left | \Phi_\beta \right )
= \left \langle \Psi_\alpha \right| \hat{V} \left | \Psi_\beta \right \rangle.
$
and
$ \left ( \Phi_\alpha \right | \hat{\cal N}_B \left | \Phi_\beta \right )
= \left \langle \Psi_\alpha \right. \left | \Psi_\beta \right \rangle$.
We term $\hat{\cal T}_B$,$\hat{\cal V}_B$ the boson {\it representations} of
the fermion operators $\hat{T}$, $\hat{V}$.  One finds the `linked-cluster' (a
la Kishimoto and Tamura\cite{KT}
although with differences) expansion of the
representations to be of the form \cite{GJ}
\begin{equation}
\hat{\cal N}_B = 1 + \sum_{\ell=2}^\infty
\sum_{\left \{ \sigma, \tau \right \} }
w_\ell^0( \sigma_1, \ldots, \sigma_l; \tau_1, \ldots, \tau_l)
\prod_{i=1}^\ell  b^\dagger_{\sigma_i}
\prod_{j=1}^\ell  b_{\tau_j} .
\label{NormExpand}
\end{equation}
and similarly for $\hat{\cal V}_B, \hat{\cal T}_B$.  In the norm operator
the $\ell$-body terms embody the
fact that the fermion-pair operators do not have exactly bosonic
commutation relations, and act to enforce the Pauli principle.
The coefficients $w_\ell^0$ etc.\ can be written in closed, albeit
complicated,   form \cite{GJ}.

The norm operator  can be  conveniently and compactly
expressed \cite{Doba,GJ,JG} in terms of the  $k$th order Casimir operators of
the unitary group ${\rm SU}(2\Omega)$, $\hat{C}_k = 2^k \,{\rm tr \,} \left(
{\bf P }
\right)^k$, ${\bf P }=
\sum_{\sigma \tau}
{b}^\dagger_\sigma {b}_\tau {\bf A}_\sigma {\bf A}^\dagger_\tau$ (and so is
both a matrix and a boson operator; the trace is over the matrix indices
and not the boson Fock space)
\begin{equation}
\hat{\cal N}_B
= \, \colon \exp \left ( - { 1\over 2}
\sum_{k=2}^\infty
{ (-1)^{k} \over k}\hat{C}_k
\ \right ) \colon
\end{equation}
where the colons `:' refer to normal-ordering of the boson operators.
This norm operator takes into account the exchange terms in the BZ expansion
of a fermion pair given in (\ref{BZpair}).

Similarly --- and this is a new result we have not seen elsewhere in
the literature  --- the
representations $\hat{\cal T}_B, \hat{\cal V}_B$ can also be written
in compact form\cite{GJ,JG}:
\begin{eqnarray}
\hat{\cal T \rm}_B & = & 2 \sum_{\sigma,\tau} \colon {\rm tr \,
}
\left[ {\bf A}_\sigma {\bf T} {\bf A}^\dagger_\tau {\bf G} \right]
b^\dagger_\sigma b_\tau \hat{ \cal N}_B \colon
\label{OneBodyImage}
\\
\hat{\cal V}_B & = &
\sum_{\mu, \nu}
\left \langle \mu \left | V \right | \nu \right \rangle
\sum_{\sigma, \tau} \colon \left \{
{\rm tr \,} \left [ {\bf A}_\sigma {\bf A}^\dagger_\mu {\bf G} \right ]
{\rm tr \,} \left [ {\bf A}_\nu {\bf A}^\dagger_\tau {\bf G} \right ]
\right. \nonumber \\
\hspace*{1.00in} & + & \left.4 \,{\rm tr \,} \left [
{\bf A}_\sigma {\bf A}^\dagger_\mu {\bf P G A}_\mu {\bf A}^\dagger_\tau
{\bf G} \right ] \right \}
b^\dagger_\sigma b_\tau  \hat{ \cal N}_B \colon ,
\label{TwoBodyImage}
\end{eqnarray}
where ${\bf G} = ({\bf 1} + 2{ \bf P})^{-1}$.
These compact forms are useful for formal manipulation.  Furthermore
they have  the powerful property of exactly expressing the
fermion matrix elements under {\it any} truncation, a fact not previously
appreciated in the literature even for the norm operator \cite{Doba}.
By this we mean the following: suppose we truncate our fermion Fock space
to states constructed from a restricted set of pairs
$\left \{ \bar{\sigma } \right \}$. Such a truncation need {\it not} correspond
to any subalgebra.   Then the representations in the corresponding
truncated boson space, which still exactly reproduce the fermion matrix
elements and which we denote by $\left [ {\cal N }_B \right ]_T$ etc.,
are  the same as those given above, retaining only the `allowed'
bosons with unrenormalized coefficients.  For example
\begin{equation}
\left[{\cal N \rm}_B \right]_T = \colon \exp \left(-\frac{1}{2}
\sum^{\infty}_{k=2} \frac{(-1)^k}{k} \left[\hat{C}_k\right]_T\right):
\end{equation}
where
\begin{equation}
[\hat{C}_k]_T = 2  \colon {\rm tr \, } (\left [{\bf P}\right ]_T )^k
 \colon \, ,\, \left [{\bf P}\right ]_T =  \sum_{\bar{\sigma}\bar{\tau}}
b^{\dagger}_{\bar{\sigma}} b_{\bar{\tau}} {\bf
A}_{\bar{\sigma}} {\bf A}^{\dagger}_{\bar{\tau}}.
\end{equation}
This invariance of the coefficients under truncation will
not hold true for the boson {\it images} introduced below.

With the boson representations of fermion operators in hand, one can
express the fermion Schr\"odinger equation (\ref{Schrodinger1})
 with $\hat{H} = \hat{T} +
\hat{V}$ as a generalized boson eigenvalue equation,
\begin{equation}
\hat{\cal H}_B \left | \Phi_\lambda \right )
= E_\lambda \hat{ \cal N}_B   \left | \Phi_\lambda \right ).
\label{Schrodinger2}
\end{equation}
Here $\hat{\cal H}_B$ is the boson representation of the fermion Hamiltonian.
Every physical fermion eigenstate in (\ref{Schrodinger1})
has a corresponding eigenstate, with the same eigenvalue, in
(\ref{Schrodinger2}).
Because the space of states constructed from pairs of fermions is
overcomplete, there also exist spurious boson states that do not
correspond to unique physical fermion states.
These spurious states will have zero eigenvalues and so can be identified.
 The overcompleteness also
means that (\ref{Schrodinger2})
 is harder to solve exactly than (\ref{Schrodinger1}).
So one truncates the model
space.

\section{Boson images}

In general the boson representations given in (\ref{NormExpand}),
(\ref{OneBodyImage}) and (\ref{TwoBodyImage})
do not have good convergence properties,
so that simple termination of the series such as (\ref{NormExpand})
in $\ell$-body
terms is impossible and use of the generalized eigenvalue
equation
(\ref{Schrodinger2}), as written, is problematic.
Instead we ``divide out'' the norm operator to obtain the {\it boson image},
i.e.\ schematically,
\begin{equation}
\hat{h} \sim ``\hat{{\cal H \rm}}_B / \hat{{\cal N \rm}}_B."
\end{equation}
That this is reasonable is suggested by the  explicit forms of
(\ref{OneBodyImage}) and  (\ref{TwoBodyImage}).
The hope of course is that $h$ is finite or nearly so, so that a 1+2-body
fermion Hamiltonian is mapped to an image
\begin{equation}
\hat{h} \sim \theta_1 b^\dagger b + \theta_2  b^\dagger b^\dagger b b
+ \theta_3  b^\dagger b^\dagger b^\dagger b b b
+ \theta_4  b^\dagger b^\dagger b^\dagger b^\dagger b b b b + \ldots
\label{HTruncate}
\end{equation}
 with the
 $\ell$-body terms, $\ell > 2$,  zero or greatly suppressed.
We now discuss how this ``dividing out'' is to be carried out.

\subsection{Exact results:  Full Space}

It turns out that for a number of cases the image of the
Hamiltonian is exactly finite. In particular, for the full boson Fock space
the representations factor in a
simple way:
$\hat{\cal T}_B = \hat{\cal N}_B \hat{T}_B =  \hat{T}_B \hat{\cal N}_B$ and
$\hat{\cal V}_B = \hat{\cal N}_B \hat{V}_B =  \hat{V}_B \hat{\cal N}_B$,
where the factored operators $\hat{T}_B, \, \hat{V}_B$, which we term  the
boson images of  $\hat{T}, \, \hat{V}$, have
simple form \cite{GJ,JG}:
\begin{equation}
 \hat{T}_B =
2\sum_{\sigma \tau}
{\rm tr \,}
\left ({\bf A}_\sigma {\bf T A}_{\tau}^\dagger
 \right)
{b}^\dagger_{\sigma} {b}_\tau,
\end{equation}
\begin{equation}
\hat{V}_B  = \sum_{\mu \nu} \left \langle \mu  \right |
V \left | \nu \right \rangle \left[
{b}^\dagger_\mu {b}_\nu
+ 2 \sum_{\sigma \sigma^\prime} \sum_{\tau \tau^\prime}
{\rm tr \,}
\left (
{\bf A}_\sigma {\bf A}^\dagger_\mu
{\bf A}_{\sigma^\prime} {\bf A}^\dagger_\tau
{\bf A}_\nu {\bf A}^\dagger_{\tau^\prime}
\right )
{b}^\dagger_\sigma {b}^\dagger_{\sigma^\prime}
{b}_\tau {b}_{\tau^\prime}\right]
\end{equation}
This image Hamiltonian
$ \hat{H}_B = \hat{T}_B + \hat{V}_B$ is the one determined by BZ if one
decomposes the Hamiltonian into multipole-multipole form and then
maps these mulitpole operators.  As discussed earlier, these BZ multipole
operators are finite in the full space.
This result, and its relation to other mappings, was noted by  Marshalek
\cite{Marshalek,Review}

Thus any boson representation of a Hamiltonian factorizes:
$\hat{\cal H}_B = \hat{\cal N}_B \hat{H}_B$ in the full space.  Since the norm
operator is a
function of the ${\rm SU}(2\Omega)$ Casimir operators it commutes with the
boson
images of fermion operators \cite{GJ,JG}, and one can simultaneously
diagonalize both $\hat{\cal H}_B$ and $\hat{\cal N}_B$.  Then
Eqn.\ (\ref{Schrodinger2}) becomes
\begin{equation}
 \hat{H}_B\left | \Phi_\lambda \right ) =
E^\prime_\lambda \left | \Phi_\lambda \right ).
\label{Schrodinger3}
\end{equation}
where $E^\prime_\lambda = E_\lambda$
for the physical states, but $E^\prime_\lambda$ for the
spurious states is no longer necessarily zero.
The boson Hamiltonian $\hat{H}_B$ is by construction Hermitian and, if one
starts with at most only two-body interactions between fermions, has at most
two-body
boson interactions.  All physical eigenstates of the original fermion
Hamiltonian will have counterparts
in (\ref{Schrodinger3}).  It should  be clear that transition amplitudes
between physical eigenstates will be preserved. Spurious states will also
exist but, since the norm operator $\hat{\cal N}_B$ commutes with the boson
image
Hamiltonian $\hat{H}_B$, the physical eigenstates and the spurious states
will not admix.  Also the spurious states can be identified because, while
they will no longer have zero energy eigenvalues, they will have eigenvalue
zero with respect to the norm operator.

\subsection{Exact Results:  Truncated space}

The boson Schr\"odinger equation (\ref{Schrodinger3}),
though finite, is not
much use as the boson Fock space is still much larger than the original
fermion Fock space, and we still must truncate the boson Fock space.
Although the representations remain exact under truncation,
the factorization  into the image does not persist in general:
 $\left [ \hat{\cal H}_B \right ]_T
\neq   \left [ \hat{\cal N}_B \right ]_T \left [ \hat{H}_B \right ]_T$.
This was recognized by Marshalek \cite{Marshalek}.
(An alternate formulation \cite{Marshalek} does not require the
complete Fock space, but mixes physical and spurious states and so always
requires a projection operator.)

If the  truncation scheme represents a closed subalgebra (specifically,
if the truncated set of fermion pairs are
closed under double commutations)  then a factorization \cite{JG2}
\begin{equation}
\left [ \hat{\cal H}_B \right ]_T
=  \left [ \hat{\cal N}_B \right ]_T \hat{h}_D
=   \hat{h}_D^\dagger \left [ \hat{\cal N}_B \right ]_T
\label{DysonFactor}
\end{equation}
{\it does} exist, with $\hat{h}_D$ at most two-body, but not
necessarily Hermitian.  We term it a {\em Dyson} image \cite{Dyson,Review}.
Under more restricted  conditions on  the structure of the pairs and
the Hamiltonian one can guarantee $\hat{h}_D$ is Hermitian and commutes with
$ \left [ \hat{\cal N}_B \right ]_T$.  In the full space, of course,
all definitions of boson images coincide and yield the same result.

First, consider a partition of the single fermion states labeled by $i = (i_a,
i_c)$, where the dimension of each subspace is $2\Omega_a$, $2\Omega_c$  so
that $\Omega = 2 \Omega_a \Omega_c$.  We denote the amplitudes for the
truncated space as ${\bf A}^{\dagger}_{\bar{\alpha}}$ and assume they can be
factored, $ ({\bf A}^{\dagger}_{\bar{\alpha}})_{ij} = ({\bf K}^{\dagger})_{i_a
j_a} \otimes (\bar{{\bf A}}^{\dagger}_{\bar{\alpha}})_{i_c j_c}$, with
${\bf K}^{\dagger} {\bf K} = {\bf K K}^{\dagger} = \frac{1}{2 \Omega_a}$ and
${{\bf K}^T} = (-1)^p {\bf K}$, where $p = 0$ (symmetric) or $p = 1$
(antisymmetric).

Furthermore we assume the completeness relation (\ref{completeness}),
which was crucial for proving
that $\hat{\cal H}_B = \hat{\cal N}_B \hat{H}_B$ \cite{GJ,JG},
is valid for the  truncated space; i.e.,
\begin{equation}
\sum_{\bar{\alpha}} (\bar{\bf A}^{\dagger}_{\bar{\alpha}})_{i_c j_c}
(\bar{\bf A}_{\bar{\alpha}})_{j_c^{\prime} i_c^\prime}
= \frac{1}{2} \left[ \delta_{i_c, i^{\prime}_c} \delta_{j_c ,j^{\prime}_c} -
(-1)^p \delta_{i_c, j^{\prime}_c} \delta_{i^{\prime}_c,j_c} \right].
\label{newcomplete}
\end{equation}
The norm operator in the truncated space then becomes
\begin{equation}
\left[\hat{\cal{N}}_B\right]_T  = \colon {\rm exp} \sum_{k = 2}
\left(\frac{-1}{\Omega_a}\right)^{k-1} \frac{1}{k}
{\rm tr} (\bar{\bf P}^k) \colon ,
\end{equation}
where $\bar{\bf P} = \sum_{\bar{\sigma} \bar{\tau}} b^{\dagger}_{\bar{\sigma}}
b_{\bar{\tau}} \bar{\bf A}_{\bar{\sigma}}
\bar{\bf A}^{\dagger}_{\bar{\tau}}$ so that
 $\left[{ \bf P}\right]_T = \left(\frac{1}{2
\Omega_{a}}\right)\bar{\bf P}$.
In this case the boson image of a one-body operator is the truncation of the
boson image in the full space,
\begin{equation}
\left [ \hat{\cal T}_B \right ]_T =
\left [ \hat{\cal N}_B \right ]_T
\left [ \hat{ T}_B \right ]_T
\end{equation}
\begin{equation}
\left [ \hat{ T}_B \right ]_T =
2\sum_{ \bar{\sigma}, \bar{\tau}}
{\rm tr \,} \left ( {\bf A}_{\bar{\sigma}} {\bf T}
{\bf A}^\dagger_{\bar{\tau}} \right )
b^\dagger_{\bar{\sigma}} b_{\bar{\tau}}.
\end{equation}
The representation of a two-body interaction can be factored into a boson image
 times the truncated norm,
\begin{equation}
\left [ \hat{\cal V}_B \right ]_T =
\left [ \hat{\cal N}_B \right ]_T \hat{v}_D;
\label{DysonImage2}
\end{equation}
however, $\hat{v}_D$, while finite (1+2-body), is not simply related to
$\left [ \hat{V}_B \right ]_T$ as is the case for one-body operators. If one
writes
\begin{equation}
\hat{v}_D =
\sum_{\bar{\sigma}, \bar{\tau}} \langle \bar{\sigma}|V|\bar{\tau} \rangle
b^{\dagger}_{\bar{\sigma}} b_{\bar{\tau}} +
\sum_{\bar{\sigma}\bar{\sigma}^{\prime}\bar{\tau}\bar{\tau}^{\prime}}\langle
\bar{\sigma} \bar{\sigma}^{\prime} |v| \bar{\tau} \bar{\tau}^{\prime} \rangle
b^{\dagger}_{\bar{\sigma}} b^{\dagger}_{\bar{\sigma}^{\prime}} b_{\bar{\tau}}
b_{\bar{\tau}^{\prime}},
\end{equation}
then matrix elements of the two-boson interaction
are
\begin{eqnarray}
\left \langle \bar{\sigma} \bar{\sigma}^{\prime} | v | \bar{\tau}
\bar{\tau}^{\prime} \right \rangle =
\nonumber \\
\frac{\sum \left \langle \mu | V | \nu
\right \rangle}{\Omega_a (2 \Omega_a - (-1)^p)(\Omega_a + (-1)^p)} {\rm tr}_a
\lbrace {\rm tr}_c (\bar{\bf A}_{\bar{\sigma}} \bar{{\bf A}}^\dagger_{\tau}
{\bf A}_{\bar{\nu}} \bar{\bf A}^\dagger_{\bar{\tau}^{\prime}}) {\rm tr}_c
(\bar{\bf A}_{\bar{\sigma}^\prime} {\bf A}^\dagger_{\mu})
\nonumber \\
+ 2 \Omega_{a} [{\rm tr}_c (\bar{\bf A}_{\bar{\sigma}} {\bf A}^\dagger_{\mu}
\bar{\bf A}_{\bar{\sigma}^{\prime}} \bar{\bf A}^\dagger_{\bar{\tau}} {\bf
A}_{\nu} \bar{\bf A}^\dagger_{\bar{\tau}^{\prime}}) - {\rm tr}_c (\bar{\bf
A}_{\bar{\sigma}} \bar{\bf A}^\dagger_{\mu} \bar{\bf A}_{\bar{\sigma}^{\prime}}
 \bar{\bf A}^\dagger_{\bar{\tau}} {\bf A}_{\nu} {\bf K} \bar{\bf
A}^\dagger_{\bar{\tau}^{\prime}})]
\nonumber
\\
- \Omega_a (2\Omega_a + (-1)^p) {\rm tr}_c ({\bf A}_{\nu} {\bf K} \bar{\bf
A}^\dagger_{\bar{\tau}} \bar{\bf A}_{\bar{\sigma}^{\prime}} \bar{\bf
A}^\dagger_{\bar{\tau}^{\prime}}) \delta_{\bar{\sigma}, \mu} \rbrace .
\end{eqnarray}
Upon inspection one sees the image is not constrained to be Hermitian.

Consider the additional condition between the matrix elements of
the interaction:
\begin{eqnarray}
\sum_{\mu, \nu} \left \langle \mu \left | V \right | \nu \right \rangle
\sum_{i_a, j_a}
\left (  {\bf A}_\nu \right )_{ i_a i_c, j_a j_c}
\left ( {\bf A}^\dagger_\mu  \right )_{j_a j_c^\prime, i_a i_c^\prime}
\nonumber \\
= N_a
\sum_{\mu, \nu}
 \left \langle \mu \left | V \right | \nu \right \rangle
\sum_{i_a, j_a}
\left (  {\bf A}_\nu \right )_{ i_a i_c, j_a j_c}
\left ( {\bf K}^\dagger \right ) _{ j_a, i_a}
\sum_{i_a^\prime, j_a^\prime}
\left ( {\bf K} \right ) _{ i_a^\prime, j_a^\prime}
\left ( {\bf A}^\dagger_\mu  \right )_{j_a^\prime j_c^\prime,
i_a^\prime i_c^\prime}
\label{HermCondition}
\end{eqnarray}
where the factor $N_a=\Omega_a(2\Omega_a+(-1)^p) $ is the number of pairs in
the excluded subspace.  While condition (\ref{HermCondition})
looks complicated there are
interactions that satisfy it; for example, two-body interactions constructed
from one-body operators $\hat{V} = \hat{T}_{\bar{\alpha} \bar{\beta}}
\hat{T}_{\bar{\alpha}^\prime \bar{\beta}^\prime}$ where $\hat{T}_{\bar{\alpha}
\bar{\beta}} = \left [A^\dagger_{\bar{\alpha}} , A_{\bar{\beta}} \right ]$.
When (\ref{HermCondition})
is satisfied then $\hat{v}_D$ is Hermitian and although $\hat{v}_D
\neq \left [\hat{V}_B \right ]_T$ they are simply related:
\begin{eqnarray}
\hat{v}_D = \sum_{\bar{\sigma},\bar{\tau}}
\left \langle \bar{\sigma} \left | V \right | \bar{\tau} \right \rangle
b^\dagger_{\bar{\sigma}} b_{\bar{\tau}}
\nonumber \\
 + 2 f_{\Omega_a}
\sum_{{\mu},{\nu}} \left \langle {\mu} \left | V \right | {\nu} \right \rangle
\sum_{  \bar{\sigma} \bar{\sigma}^\prime,
\bar{\tau} \bar{\tau}^\prime }
{\rm tr \,} \left (
{\bf A}_{\bar{\sigma}} {\bf A}^\dagger_{{\mu}}
{\bf A}_{\bar{\sigma}^\prime} {\bf A}^\dagger_{\bar{\tau}}
{\bf A}_{{\nu}} {\bf A}^\dagger_{\bar{\tau}^\prime}
\right )
b^\dagger_{\bar{\sigma}} b^\dagger_{\bar{\sigma}^\prime}
b_{\bar{\tau}} b_{\bar{\tau}^\prime}
\label{HermDysonImage}
\end{eqnarray}
with  $f_{\Omega_a} = 4\Omega_a^2/N_a $ renormalizing the two-boson part of $
\left [ \hat{V}_B \right ]_T$ by a factor  which
ranges from unity (full space) to 2 for a very small subspace.
Not all interactions satisfy (\ref{HermCondition});
for example, the pairing interaction never
does except in the full space.  For the pairing interaction
$\left \langle  {\mu} \left |
V^{\rm pairing} \right | {\nu} \right \rangle = \delta_{\mu, 0}
\delta_{\nu, 0}G$, and
${\bf A}_{0} {\bf A}_0^\dagger = \frac{1}{2 \Omega}$,
and the image (\ref{DysonImage2}) $\hat{v}_{D}^{\rm pairing} $ becomes
(remembering $\Omega = 2 \Omega_a \Omega_c$)
\begin{equation}
 G \left \lbrace \hat{{\rm N}}_0 [ 1 - \frac{2}{\Omega}
\hat{{\rm N}} + \frac{1}{\Omega} + \frac{\hat{{\rm N}}_0}{\Omega}]
 + \sum_{\bar{\tau} \bar{\tau}^\prime \neq 0, \bar{\sigma}} {\rm tr}\;
(\bar{{\bf A}}_{\bar{\sigma}} \bar{{\bf A}}^\dagger_{\bar{\tau}} \bar{{\bf
A}}_0 \bar{{\bf A}}_{\bar{\tau}^\prime}^\dagger) b^\dagger_{\bar{\sigma}}
b^\dagger_0 b_{\bar{\tau}} b_{\bar{\tau}^\prime} \right \rbrace,
\label{PairingImage}
\end{equation}
where $\hat{{\rm N}}$ is the total number of bosons, $\hat{{\rm N}} =
\sum_{\bar{\tau}} b^\dagger_{\bar{\tau}} b _{\bar{\tau}}$, and $\hat{{\rm N}}_0
=  b^\dagger_0 b _0$.  The second term in the above is not Hermitian but can be
transformed away by a similarity transformation \cite{GT},
leaving the first term as  a finite Hermitian image which gives the correct
eigenvalues for all N.

The SO(8) and Sp(6) models \cite{SO8} belong to a class of models which have a
subspace for which (\ref{newcomplete}) is valid and interactions which
satisfy (\ref{HermCondition}).
In these models the shell model orbitals
have a definite angular momentum $\vec{\bf j}$ and
are partitioned into a pseudo orbital angular momentum $\vec{\bf k}$ and
pseudospin $\vec{\bf i}$,
$\vec{\bf j} = \vec{\bf k}+ \vec{\bf i}$.  The
amplitudes are then given as products of Clebsch-Gordon coefficients,
\begin{equation}
\left(A^{\dagger}_{\alpha}\right)_{ij} = \frac{(1 + (-1)^{K + I})}{2}
(k \, m_i, k\, m_j |K_\alpha \, M_\alpha) \, (i \, \mu_i, i\, \mu_j |I_\alpha
\, \mu_\alpha),
\end{equation}
 where $K$ and $I$ are the total pseudo orbital angular
momentum and pseudospin respectively of the pair of nucleons.  For the SO(8)
model  ${\bf i} = \frac{3}{2}$ and one  considers the subspace of pairs with $K
 = 0$ $(p = 0)$, $(\bar{A}^{\dagger}_{\bar{\alpha}})_{ij} = \frac{(1 +
(-1)^I)}{2} ( i \, \mu_i, i \, \mu_j | I_\alpha \, \mu_\alpha)$;
in the Sp(6) model  ${\bf k} = 1$ and one considers the subspace with $I = 0$
$(p = 1)$, $(\bar{A}^{\dagger}_{\bar{\alpha}})_{ij} = \frac{(1 + (-1)^K)}{2}
(k \, m_i, k\, m_j | K_\alpha \, M_\alpha)$.
The complicated conditions (\ref{HermCondition})
hold true for important cases, for example, the
quadrupole-quadrupole and other multipole-multipole interactions in the
SO(8) and Sp(6) models (that is, interactions of the generic form
$P^r \cdot P^r$ in the notation of \cite{SO8})
have Hermitian Dyson images.  Not all interactions in these models
have Hermitian Dyson images.  For example, pairing in any model
(see (\ref{PairingImage}) )and,
in the SO(8) model, the particular
combination $g_0( S^\dagger S + {1\over 4} P^2\cdot P^2 )$ which is the
SO(7) limit.
It so happens that these particular cases nonetheless
can be brought into finite, Hermitian form as discussed in the
next section.

\subsection{  Approximate or numerical images \label{ApproxImages} }

The most general image Hamiltonian one can define is
\begin{equation}
\hat{h} \equiv {\cal U} \left [ \tilde{\cal N}_B \right ]_T^{-1/2}
\left [\hat{{\cal H \rm}}_B \right ]_T \left [ \tilde{\cal N}_B \right
]_T^{-1/2} {\cal U}^\dagger,
\label{HImage}
\end{equation}
which is manifestly Hermitian for any truncation scheme and any
interaction, with ${\cal U}$  a unitary operator.
(Because the norm is a singular operator
it cannot be inverted. Instead $\left [ \tilde{\cal N}_B \right ]_T^{-1/2}$
is calculated from the norm only in the physical subspace,
with the zero eigenvalues which annihilate the spurious states retained.
Then $\hat{h}$ does not mix physical and spurious states.)  If ${\cal U} = 1$,
this is  the democratic mapping \cite{democratic}. Again, for the
full space $\hat{h} = \hat{h}_D = \hat{H}_B$.

This prescription is, we argue, useful for a practical derivation of boson
image Hamiltonians.  Ignoring for the moment the unitary transformation
${\cal U}$, consider the expansion (\ref{HTruncate}) of $\hat{h}$.
The operators $\left [\hat{{\cal H \rm}}_B \right ]_T$ and
$\left [ \tilde{\cal N}_B \right ]_T^{-1/2}$ have similar expansions, and by
multiplying out (\ref{HImage})
one sees immediately that the coefficient $\theta_\ell$ depends only
on up to $\ell$-body terms in
$\left [ {\cal H}_B \right ]_T$ and
$\left [ \tilde{\cal N}_B \right ]_T^{-1/2}$, derived from $2\ell$-fermion
matrix elements which are tractable for $\ell$ small.
Ideally $\hat{h}$ would have at most two-body terms,  and
our success in finding finite images in the previous section gives us
hope that the high-order many-body terms may be small; at any
rate the convergence can be calculated and checked term-by-term.
Specifically, consider the convergence
of the series (\ref{HTruncate}) as a function of $\ell$. A rough estimate
is that, for an $N$-boson Fock space, one can truncate to the $\ell$-body
terms if for $\ell^\prime > \ell$,  $\theta_{\ell^\prime}$
is sufficiently small compared to
$\theta_\ell \times (N-\ell^\prime)!/(N-\ell)!$; the strictest condition is to
require $\theta_{\ell^\prime} \ll \theta_\ell /(\ell^\prime-\ell)!$.

The Hermitian image $\hat{h}$, defined in (\ref{HImage}),
is related to the Dyson image  $\hat{h}_D$, defined in (\ref{DysonFactor}),
by a similarity transformation
${\cal S} = {\cal U} \left [\tilde{\cal N}_B \right ]_T^{1/2}$,
\begin{equation}
\hat{h} = {\cal S} \hat{h}_D {\cal S}^{-1}.
\end{equation}
 The similarity transformation
${\cal S}$ orthogonalizes the fermion states
$\left | \Psi_{\bar{\alpha}} \right \rangle$  inasmuch
$({\cal S }^{-1})^\dagger\tilde{\cal N }_B{\cal S}^{-1} = 1$
in the physical space (and $=0$ in the spurious space). This
is akin to Gram-Schmidt orthogonalization and the freedom to choose ${\cal U}$,
and ${\cal S}$, corresponds to the freedom one has in ordering the
vectors to be Gram-Schmidt orthogonalized.  The OAI and democratic mappings
are just two particular choices out of many; the
latter takes ${\cal U}=1$.
We can use the freedom in the choice of ${\cal U}$ to our advantage.
Consider the SO(8) model \cite{SO8}
 and its three algebraic limits: the pure pairing
interaction, the quadrupole $P^2\cdot P^2$ interaction,
which can be written in terms of
SO(6) Casimir operators, and the linear combination of pairing and quadrupole
$S^\dagger S + {1 \over 4} P^2\cdot P^2$ which can be written in terms of SO(7)
 Casimirs (see \cite{SO8} for details and notation).
As discussed in the last Section, the Dyson image of the quadrupole interaction
is Hermitian and finite, and $\hat{h}_D = \hat{h}$ with
${\cal U} = 1$.  The Dyson images of
the pairing and SO(7) interactions are finite but non-Hermitian.  We have
found ${\cal U}$'s $\neq 1$ for both these cases
(but not the same ${\cal U}$) such that
their respective Hermitian images $\hat{h}$ are finite; the one for pairing
is exactly the OAI prescription, while that for SO(7) is exactly opposite,
orthogonalizing states of low seniority against states of higher
seniority.  These general Hermitian images do not have a simple relation
to the truncated full image, as do the Hermitian Dyson image
(\ref{HermDysonImage}).
The one-body piece remains unchanged but there can be significant
renormalization, and even change of sign, of the two-body piece.
For example, for the pairing interaction
$\left [\hat{H}_B \right ]_T = s^{\dagger} s + {1\over 2 \Omega^2}
s^{\dagger} s^{\dagger} s s+$ additional terms, including off-diagonal terms
such as $d^{\dagger} d^{\dagger} ss$, whereas $\hat{h} =  s^{\dagger} s -
{1\over \Omega}
s^{\dagger} s^{\dagger} s s+$ (depending on the truncation) terms such as
$d^{\dagger} d^{\dagger} \tilde{d} \tilde{d}$ but {\it no} off-diagonal terms.
Hence we see here the renormalization is not just a simple overall factor $f$
as it was for the Hermitian Dyson image: it is
$-2 \Omega$ for the $ s^{\dagger} s^{\dagger} s s$ term but $0$ for
terms such as $d^{\dagger} d^{\dagger} s s.$
\begin{figure}
\vspace{105mm}
\caption{ Spectrum of SO(7) interaction, for 7 bosons,
in SO(8) model with exact
(left) and approximate (right) two-body boson Hamiltonians.}
\end{figure}

If one uses an ``inappropriate'' transform ${\cal S \rm}$ it can induce
an unneeded and unwanted many-body dependence.  This principle we
illustrate  in  the SO(8) model with the SO(7) interaction, whose spectrum is
exactly known  and for which we can derive a finite Hermitian image with no
many-body dependence; this is the left-hand spectrum in Figure 1.
For the right-hand side we took the ${\cal U}$
appropriate for pairing, that is the
OAI prescription determined for $N = 2$, and calculated the spectrum for
$N=7$  keeping
only the strict two-body
terms.  The distortion in the spectrum from the exact result, such as
the overall energy shift and the large perturbation in the third band,
indicates missing many-body terms. That is, if one mapped the SO(7)
interaction using the canonical OAI procedure one would find of
necessity a many-body dependence in the interaction coefficients.  By
orthogonalizing the basis in the a different way, however, as expressed by a
different choice of ${\cal U}$, the many-body dependent vanishes.
Therefore it is possible that some of the $N$-dependence of OAI is induced by
their choice of orthogonalization and could be minimized with a
different choice.  We are currently exploring how to exploit this freedom to
best effect.

\section{ Summary}

In order to investigate rigorous foundations for the phenomenological
Interacting Boson Model,
we have presented a rigorous microscopic mapping of fermion pairs to bosons,
paying special attention to exact mapping of matrix elements, Hermiticity,
truncation of the model space, and many-body terms.
First we presented new, general and compact forms for boson
{\it representations} that preserve fermion matrix elements. We then
considered the
boson {\it image} Hamiltonian which results from ``dividing out'' the
norm from the representation; in the full boson Fock
space the image is always finite and Hermitian; in addition we discussed
several analytic cases for truncated spaces where the image is also
finite and Hermitian.  Finally, we give a prescription
which is a generalization of both the OAI and democratic mappings;
in the most general case for  truncated spaces the Hermitian image
Hamiltonian may not be finite but we have demonstrated there is some
freedom in the mapping that one could possibly exploit to minimize the
many-body terms. This freedom, which manifests itself in a similarity
transformation that orders the orthogonalization of the underlying fermion
basis, depends on the Hamiltonian.

This research was  supported by the U.S.\ Department of Energy.
The boson calculations for Figure 1 were performed using the PHINT package of
Scholten \cite{Scholten}.

\end{document}